\documentstyle[12pt]{article}

\catcode`\@=11
\long\def\@makefntext#1{
\protect\noindent \hbox to 3.2pt {\hskip-.9pt
$^{{\ninerm\@thefnmark}}$\hfil}#1\hfill}		

 \def\@makefnmark{\hbox to 0pt{$^{\@thefnmark}$\hss}}  

\def\ps@myheadings{\let\@mkboth\@gobbletwo
\def\@oddhead{\hbox{}
\rightmark\hfil\ninerm\thepage}
\def\@oddfoot{}\def\@evenhead{\ninerm\thepage\hfil
\leftmark\hbox{}}\def\@evenfoot{}
\def\sectionmark##1{}\def\subsectionmark##1{}}

\newcounter{sectionc}\newcounter{subsectionc}\newcounter{subsubsectionc}
\renewcommand{\section}[1] {\vspace{0.6cm}\addtocounter{sectionc}{1}
\setcounter{subsectionc}{0}\setcounter{subsubsectionc}{0}\noindent
	{\bf\thesectionc. #1}\par\vspace{0.4cm}}
\renewcommand{\subsection}[1] {\vspace{0.6cm}\addtocounter{subsectionc}{1}
	\setcounter{subsubsectionc}{0}\noindent
	{\it\thesectionc.\thesubsectionc. #1}\par\vspace{0.4cm}}
\renewcommand{\subsubsection}[1]
{\vspace{0.6cm}\addtocounter{subsubsectionc}{1}
	\noindent {\rm\thesectionc.\thesubsectionc.\thesubsubsectionc.
	#1}\par\vspace{0.4cm}}

\newcounter{appendixc}
\newcounter{subappendixc}[appendixc]
\newcounter{subsubappendixc}[subappendixc]

\renewcommand{\appendix}[1] {\vspace{0.6cm}
        \refstepcounter{appendixc}
        \setcounter{figure}{0}
        \setcounter{table}{0}
        \setcounter{equation}{0}
        \renewcommand{\thefigure}{\Alph{appendixc}.\arabic{figure}}
        \renewcommand{\thetable}{\Alph{appendixc}.\arabic{table}}
        \renewcommand{\theappendixc}{\Alph{appendixc}}
        \renewcommand{\theequation}{\Alph{appendixc}.\arabic{equation}}
        \noindent{\bf Appendix \theappendixc #1}\par\vspace{0.4cm}}

\def\abstracts#1{{
	\centering{\begin{minipage}{30pc}\tenrm\baselineskip=12pt\noindent
	\centerline{\tenrm ABSTRACT}\vspace{0.3cm}
	\parindent=0pt #1
	\end{minipage}}\par}}

\renewenvironment{thebibliography}[1]
	{\begin{list}{\arabic{enumi}.}
	{\usecounter{enumi}\setlength{\parsep}{0pt}
\setlength{\leftmargin 1.25cm}{\rightmargin 0pt}
	 \setlength{\itemsep}{0pt} \settowidth
	{\labelwidth}{#1.}\sloppy}}{\end{list}}

\newcounter{itemlistc}
\newcounter{romanlistc}
\newcounter{alphlistc}
\newcounter{arabiclistc}

\newcommand{\fcaption}[1]{
        \refstepcounter{figure}
        \setbox\@tempboxa = \hbox{\tenrm Fig.~\thefigure. #1}
        \ifdim \wd\@tempboxa > 6in
           {\begin{center}
        \parbox{6in}{\tenrm\baselineskip=12pt Fig.~\thefigure. #1}
            \end{center}}
        \else
             {\begin{center}
             {\tenrm Fig.~\thefigure. #1}
              \end{center}}
        \fi}

\newcommand{\tcaption}[1]{
        \refstepcounter{table}
        \setbox\@tempboxa = \hbox{\tenrm Table~\thetable. #1}
        \ifdim \wd\@tempboxa > 6in
           {\begin{center}
        \parbox{6in}{\tenrm\baselineskip=12pt Table~\thetable. #1}
            \end{center}}
        \else
             {\begin{center}
             {\tenrm Table~\thetable. #1}
              \end{center}}
        \fi}

\def\@citex[#1]#2{\if@filesw\immediate\write\@auxout
	{\string\citation{#2}}\fi
\def\@citea{}\@cite{\@for\@citeb:=#2\do
	{\@citea\def\@citea{,}\@ifundefined
	{b@\@citeb}{{\bf ?}\@warning
	{Citation `\@citeb' on page \thepage \space undefined}}
	{\csname b@\@citeb\endcsname}}}{#1}}

\newif\if@cghi
\def\cite{\@cghitrue\@ifnextchar [{\@tempswatrue
	\@citex}{\@tempswafalse\@citex[]}}
\def\citelow{\@cghifalse\@ifnextchar [{\@tempswatrue
	\@citex}{\@tempswafalse\@citex[]}}
\def\@cite#1#2{{$\null^{#1}$\if@tempswa\typeout
	{IJCGA warning: optional citation argument
	ignored: `#2'} \fi}}

\def\fnt#1#2{\footnotetext{\kern-.3em
	{$^{\mbox{\sevenrm #1}}$}{#2}}}

 1
 1
\font\twelveit=cmti10 scaled\magstep 1

\font\tenbf=cmbx10
\font\tenrm=cmr10
\font\tenit=cmti10

\font\ninerm=cmr9

\def\f{\phi}
\def\pa{\partial}
\def\d{\delta}
\def\D{\Delta}
\def\c{\chi}
\def\ee{\end{equation}}
\def\be{\begin{equation}}
\newcommand{\inv}[1]{\frac{1}{#1}}
\newcommand{\fder}[2]{\frac{\pa #1}{\pa #2}}
\newcommand{\sder}[1]{\frac{\pa^2}{\pa #1^2}}

\begin{document}

\centerline{\tenbf CLASSICAL AD QUANTUM IN COSMOLOGY}
\vspace{0.8cm}
\centerline{\tenrm J. KOWALSKI--GLIKMAN}
\baselineskip=13pt
\centerline{\tenit Institute for Theoretical Physics, University of
Wroc\l{}aw}
\baselineskip=12pt
\centerline{\tenit Pl. Maxa Born 9, 50204 Wroc\l{}aw, Poland}
\vspace{0.9cm}
\abstracts{There are several regimes in chaotic inflationary cosmology
where some part of the system is classical and some other quantum. I
describe how to deal with such systems and how to disentangltheir dynamics
 into
classical behaviour and quantum corrections. I also discuss the
conditions for quantum corrections to be small.
}

\vfil

Cosmology is one of the most exciting fields of modern high energy
physics. On the one hand, it provides the only  possible experimental ground
for testing theories of high energy particle physics beyond the energy
scale of accelerators (see other contributions to this proceedings). On
the other hand, because of the simple geometry of realistic cosmological
models, it makes it possible to test consistency of physical
theories at the energy
scales close to the Planck scale of quantum gravity.

 The remarkable progress in our understanding of early cosmology (by
 which I understand the models describing the universe at the energies
 between $10^{15}$ and $10^{19}$GeV) resulted in wide recognition
 of the chaotic inflationary model\cite{Linde} as the (without doubts) the most
 simple and (possibly) the only realistic model of the early evolution
 of the universe.

The basic features of the chaotic inflation are the following (see the
figure)

\begin{enumerate}
\item The universe emerges from the ``quantum gravity abyss'' with the
scale factor and the energy density of order 1.\footnote{In what follows
I use the Planck units i.e., the energy unit is $10^{19}$GeV, the length
unit is $10^{-33}$cm etc.}
\item The initial conditions for the inflaton field are such that
$\ddot{\f} \ll V'(\f)$, $(\pa_i\f)^2\ll 1$, $V(\f)\sim 1$. For the
simple massive scalar model with $m^2\sim 10^{-12}$ (as required to
secure the correct magnitude of the spectrum for large scale structures
formation), this means that the initial value of the field $\f$, $\f_0
\sim 10^6$, and $\f$ doesn't vary too much in space and time.
\item Subsequently, the field $\f$ rolls down the potential well, and the
universe is undergoing inflation ($a(t)\sim e^{Ht}$, with $H$ slowly
varying function of time.)
\item The inflation ends at the point $\f\sim 1$ and is followed by
very rapid oscillations around the minimum $\f=0$, which are interpreted
as giving rise to reheating leading to creation of  thermal hot
matter in the universe (hot big bang.)
\end{enumerate}


Within the inflationary regime, quantum fluctuations emerge superimposed
on the classical motion down the potential well. In the interval
$1\leq\f\leq 10^3$ these fluctuations are small
\be
\D_Q\f< \d_{class}\f,
\ee
where $\D_Q\f$ is the effective magnitude of the quantum fluctuations
which cummulate during some natural time interval, and $\d_{class}\f$
is the classical distance covered by the field $\f$ during the same
time. These fluctuations give rise to the seeds for large scale
formation in the standard way.

However, for $10^3\leq\f\leq 10^6$
\be
\D_Q\f> \d_{class}\f,
\ee
and the quantum fluctuations dominate the evolution of the inflaton
field, giving rise to the extremely interesting global structure of the
universe (see the paper\cite{LLM} and the references therein.)

There are therefore at least three periods during the early evolution of
the inflaton field when there is a significant interplay between
classical and quantum behaviour of the constituents of the system:

\begin{enumerate}
\item Emergence of the universe from the quantum gravity era. It is
believed that at the Planck energy scale gravity becomes classical,
however the inflaton field may be still very much quantum. How this
process looks in details? More importantly, are the initial conditions
for the chaotic inflation listed above likely or not?
\item In the subsequent inflationary phase some ``part'' of the inflaton
field is ``classical'' and some part is ``quantum.'' How the detailed
evolution of $\f$ looks like?
\item At the end of inflation the universe enters the reheating phase
with the classical oscillating inflaton and quantum radiation field.
What are details of this process?\footnote{Recently there is a renewed
intererst in analyzing the reheating process. See\cite{KLS} and references
therein.}
\end{enumerate}

This  kind of problems has been analyzed in the recent papers\cite{JV}
and\cite{V}. Following these papers, let me discuss the quantum system
consisting of the scale factor $a$  and the inflaton field $\f$. The
quantum
hamiltonian of this system is the one-dimensional Wheeler--De~Witt
operator which, at the same time, is a constraint of the theory. For the wave
function of the universe we have the equation
\be
\left[
\fder{}a \inv a \fder{}a - \inv{a^3}\sder{\f} +V(\f)a^3\right]
\Psi(\f,a)=0.                      \label{WDW}
\ee
This equation comrises the whole information obout quantum universe.
Observe that there is no time in this equation, and this is a general
feature of all theories with general coordinate invariance. (For
comprehensive discussion of the time problem in classical and quantum
gravity see the article\cite{I}.) Thus the wave function of the universe
provides
us with correlations between the scale factor and the inflaton field.
This representation is not convenient for two reasons. First, we would
like to be able to trace the time evolution of the universe, and second
we would like to see how the classical inflationary universe
emerges from the quantum gravity state. Actually, we have a very good idea of
what the classical behaviour is. This is just the classical dynamics of the
system. Therefore, should we be able to decompose the whole dynamics of the
system into
$$
(classical\;\; motion) = (corrections),
$$
we would know that the $(corrections)$ are of the quantum origin.

To solve these problems we proceed as follows. First we write the
wave function in the form
\be
\Psi(a,\f) = R(a)e^{iS(a)}\c(a,\f),
\ee
with real $R$ and $S$. This representation can be made unambiguous by
imposing certain conditions on $\c$. Substituting this into
Eq.~\ref{WDW} we find two equations, one of whose has the form
\be
-i
\inv a \fder{S}a\fder{\c}a - H_M = \ldots,
\ee
where $H_M=\inv{a^3}\sder{\f} +V(\f)a^3$ is the matter part of the
hamiltonian, where ``\ldots''  denote other correction terms. It is clear that
one can recover the standard Schr\"odinger equation (with correction
terms) for $\c$, if one introduces the new variable $t(a)$ such that
\be
\inv a \fder{S}a\fder{\c}a \stackrel{def}{=} \fder{\c}t,
\label{wf}
\ee
that is, if
\be
\fder Sa=-a\dot a,\;\;\;\; \dot a = \frac{da}{dt}. \label{time}
\ee
Then it can be shown that Eq.~\ref{WDW} is {\em strictly equivalent}
to the following set of equations
\be
a{\dot a}^2 =<H_M>+(Q_G+C_E),\label{1}
\ee
\be
i\dot{\c} -H_M\c= C_S,\label{2}
\ee
where $<H_M>$ is the quantum average of $H_M$ with respect to
$\c$ and $Q_G$, $C_E$, $C_S$ are various correction terms corresponding
to gravitational quantum correction to Einstein equations, matter
quantum corrections to Einstein equations, and matter quantum
corrections to the Schr\"odinger equation, respectively. The exact form
of these correction terms can be found in\cite{JV}.

Let me pause for a moment to discuss the logic which led us
to Eqs.~\ref{1}, \ref{2}.  Starting from Eq.~\ref{WDW}, we represented
the wave function in the form Eq.~\ref{wf}, recovered time with the
help of Eq.~\ref{time}, and ended up with the set of equations
Eq.~\ref{1}, \ref{2}. Let me stress once again  that no approximations
have been made during this procedure, Eqs.~\ref{1}, \ref{2} are
completely equivalent to the initial Eq.~\ref{WDW}. What we have now is
the system  of two equations: the Einstein equation and the
Schr\"odinger equation supplemented by quantum correction terms. This
system is clearly convenient for our purposes. In the regime where
gravity is almost classical and the inflaton field is still in the
quantum regime, we can easily see what the corrections are and under
which conditions are they small.

So what is the form and relative magnitude of the correction terms?
Typically, the quantum gravity corrections $Q_G$ and the terms $C_S$ are
proportional to $a^{-3}$, so they are getting rapidly diluted in the
inflationary phase (and in the Friedmann phase as well.) Further,
the $C_E$ term is proportional to the energy
dispersion, so it is small if quantum fluctuations of the energy
are small as compared to the energy itself. It should be noted however
that if these fluctuations are not small, there is a huge contribution
from these fluctuations to Einstein equations. One expects therefore
that if these fluctuations are not small initially, they would prevent
the emergence of the classical universe. The next point which should be
emphasized is that in  the derivation I implicitely assumed that the
wave function $\Psi$ is known. This wave function is hard to find, of
course, though some approximate solutions are being currently under
investigation\cite{BK}. On the other hand, one can, for example,
neglect all the correction terms, solve the resulting equations,
and then check explicitly if for the particular solution the correction terms
are small indeed, and then treat them as perturbations.

The procedure leading to Eqs.~\ref{1}, \ref{2} is very general, of
course, and can be applied to many systems. For example, we might be
interested in the question concerning how the evolution of the inflaton
field during inflation
splits into classical and quantum part. To do that, we
first neglect the correction term $C_S$, which is justified by the fact
that during inflation this term is rapidly approaching zero. Then we
make the polar decomposition of  $\c$
$$
\c(\f,t)=\rho(\f,t)e^{i\sigma(\f,t)}
$$
and substitute the result into Eq.~\ref{1} with $C_S=0$. The procedure
is described in\cite{JV}, the result is
\be
\ddot{\f} +3\frac{\dot a}{a}\dot{\f} +\fder{V_M}{\f} =-\fder{Q_M}{\f}
,\label{3}
\ee
where the quantity $Q_M$ encodes all quantum correction and is called
the quantum potential. We ended up therefore with the purely classical
motion with the only difference that the classical potential $V(\f)$
is now supplemented by the quantum counterpart $Q_M$. Thus the evolution
of the system is described by purely classical trajectories (in the modified
potential, though.)
One may wonder what happened to the probabilistic picture of quantum
mechanics. This part happens to be hidden in the initial conditions
to our evolutionary equations. It turns out that there is a well defined
probability measure on the set of these conditions (given by the initial wave
function), but once we pick up
a single one, the subsequent evolution is completely
probabilistic.\footnote{For the careful discussion of this point,
see\cite{V}.}

One may wonder if it is necessary to go through the intermediate step,
Eq.~\ref{2}, to obtain Eq.~\ref{3}. In fact, as it was shown in\cite{V},
one can obtain the set of equations
\be
\inv 2 a{\dot a}^2 =a^3\left(\frac{\dot{\f}}{2} +V_M(\f)\right) + Q\label{4}
\ee
\be
\ddot{\f} +3\frac{\dot a}{a}\dot{\f} +\fder{V_M}{\f} =-\fder{Q}{\f}
,\label{5}
\ee
directly from  Eq.~\ref{WDW}, without making any approximations. This means
that we are able to rewrite the
equation governing the quantum universe in the equivalent, but much more
convenient form, in which the analysis can be made in terms of
classical objects and interpretation is much easier.

It is worth mentioning that the procedure described above can be readily
extended to cover any bosonic quantum field theory. For example, in the
forthcoming
paper\cite{Jur}, this procedure is applied to the case of the quantum
field theory of massive scalar field on the De~Sitter background. It could be
also applied to the case of Yang-Mills theory and full quantum theory of
gravity.

There is one more interesting area where the above procedure can be
easily applied. In the reheating phase, we have to do with the classical
oscillating inflaton field $\f$ and the thermal field $\rho$. We again
encounter the system with the classical part (inflaton) and the quantum
subsystem (radiation). Here again one starts with the hamiltonian
$$
H = H_{\f} +H_{\rho} + H_{int}
$$
and proceeds through the steps similar to those indicated above.
Then one can proceed with the analysis of various posible outputs of
reheating.

It is interesting to note in this context that assuming that the
interaction hamiltonian is of the form $\f^2 H_I(\rho,t)$, the effective
equation for the inflaton field $\f$ the leading term looks as follows
\be
\inv 2\dot{\f}^2 + \inv 2 m^2\f^2 = < H_I>\f^2.
\ee
Above $<H_I>$ is an average computed with respect to the radiation
field, and $m$ is the effective mass of the inflaton field. The form
of this equation suggests the exciting possibility of
creation of the universe in laboratory. By appropriately preparing
the radiation field, one may make $<H_I>$ into the source for parametric
resonance for $\f$ in order to efficiently pump energy into
this field. It is worth observing that once the field $\f$ reaches the
magnitude of order one, and provided that it is sufficiently homogeneous
in space, the inflation starts and the region starts expanding rapidly.
It happens also the this region covers itself by the horizon so it, in
fact, becomes the new universe effectively disconected from our own.

Even thogh the creation of the universe is clearly impossible
using modern technology, the future experimental high energy
physicist preparing this experiment should not forget to place the
announcement ``We appologise for all the inconveniences''\cite{Adams}
somewhere in the newly created universe.
\newline


{\bf References}

\begin{thebibliography}{9}
\bibitem{Linde} A.D. Linde, {\it Phys.
Lett.} {\bf 129B} (1983) 177; see also A.D. Linde, {\it Particle Physics
and Inflationary Cosmology} (Harwood, Chur, 1990) and J.
Kowalski-Glikman {\it Eight Lectures on Cosmology} (Wroc\l{}aw
University Preprint, 1994).
\bibitem{LLM} A.D. Linde, D.A. Linde, and A. Mezhlumian, {\it
Phys. Rev.} {\bf D49} (1994) 1783.
\bibitem{KLS} L. Kofman, A.D. Linde, and A.A. Starobinsky, {\it
Reheating After Inflation} hep-th 9405187.
\bibitem{JV} J. Kowalski-Glikman and J. Vink, {\it Class. Quant. Grav.}
{\bf 7} (1990) 701.
\bibitem{V} J. Vink,  {\it Nucl. Phys} {\bf B369} (1992) 707.
\bibitem{I} C.J. Isham, {\it Canonical Gravity and the Problem of Time}
in {\it Integrable Systems, Quantum Groups, Field Theories}, L.A. Ibort
and M.A. Rodriguez Eds., (Kluwer Academic Publishers,
Dordrecht/Boston/London, 1993).
\bibitem{BK} A. B\l{}aut and J. Kowalski-Glikman, in preparation.
\bibitem{Jur} J. Kowalski-Glikman, in preparation.
\bibitem{Adams} D. Adams, {\it So Long and Thanks for All the Fish}.
\end{thebibliography}

\begin{thebibliography}{9}
\bibitem{Don/Ho} J. F. Donoghue and B. R. Holstein, {\it Phys.
Rev.} {\bf D25} (1982) 2015.
\bibitem{Coh/An} M. L. Cohen and P. W. Anderson, in {\it
Superconductivity in d- and f-Band Metals}, ed. D. H. Douglas
(AIP, New York, 1972).
\bibitem{Kre} H. Krebs, {\it Fundamentals of Inorganic Crystal
Chemistry} (McGraw-Hill, London, 1968), p. 160.
\end{thebibliography}
\end{document}